\documentclass{article}
\pdfoutput=1 
\PassOptionsToPackage{numbers}{natbib}
\usepackage[final]{nips_2017_arXiv}

\usepackage[utf8]{inputenc} 
\usepackage[T1]{fontenc}    
\usepackage{hyperref}       
\usepackage{url}            
\usepackage{booktabs}       
\usepackage{amsfonts}       
\usepackage{nicefrac}       
\usepackage{microtype}      
\usepackage{amsmath}
\usepackage{amstext}
\usepackage{amssymb}
\usepackage{graphicx}

\newcommand{\xx}{\mathbf{x}}
\newcommand{\mc}{\boldsymbol{\lambda}}

\newcommand{\Mc}{\mathbf{s}}

\newcommand{\yy}{\mathbf{y}}
\newcommand{\rr}{\mathbf{r}}

\newcommand{\VV}{\mathbf{V}}
\newcommand{\WW}{\mathbf{W}}
\newcommand{\FF}{\mathbf{F}}
\newcommand{\II}{\mathbf{I}}
\newcommand{\GG}{\mathbf{G}}
\newcommand{\DD}{\mathbf{D}}
\newcommand{\AAA}{\mathbf{A}}
\newcommand{\la}{\boldsymbol{\lambda}}
\newcommand{\La}{\boldsymbol{\xi}}
\newcommand{\Mu}{\boldsymbol{\mu}}
\newcommand{\mm}{\boldsymbol{\mu}}

\newcommand{\MAP}{\text{MAP}}
\newcommand{\Lag}{\mathcal{L}}
\newcommand{\Lolf}{\mathcal{L}_{\text{olf}}}
\newcommand{\Lmap}{\mathcal{L}_{\text{MAP}}}
\newcommand{\Lsis}{\mathcal{L}_{\text{sis}}}
\newcommand{\Lsiseps}{\mathcal{L}_{\text{sis}}^{\varepsilon}}
\newcommand{\bb}{\mathbf}
\newcommand{\zz}{\mathbf{z}}
\newcommand{\vv}{\mathbf{v}}
\newcommand{\uu}{\mathbf{u}}
\newcommand{\PP}{\mathbf{P}}
\newcommand{\diag}{\text{diag}}

\DeclareMathOperator*{\argmin}{argmin}

\title{Sparse connectivity for MAP inference in\\ linear models 
  using sister mitral cells}

\author{
  Sina Tootoonian, Peter Latham \\
  Gatsby Computational Neuroscience Unit\\
  University College London\\
  London W1T 4JG, UK \\
  \texttt{[sina|pel]@gatsby.ucl.ac.uk} \\
}

\begin{document}
\maketitle
\begin{abstract}
Sensory processing is hard because the variables of interest are encoded in spike trains in a relatively complex way. A major goal in sensory processing is to understand how the brain extracts those variables. Here we revisit a common encoding model \citep{olshausen_emergence_1996} in which variables are encoded linearly. Although there are typically more variables than neurons, this problem is still solvable because only a small number of variables appear at any one time (sparse prior). However, previous solutions usually require all-to-all connectivity, inconsistent with the sparse connectivity seen in the brain. Here we propose a principled algorithm that provably reaches the MAP inference solution but using sparse connectivity. Our algorithm is inspired by the mouse olfactory bulb, but our approach is general enough to apply to other modalities; in addition, it should be possible to extend it to nonlinear encoding models.   
\end{abstract}

\section{Introduction}
A prevalent idea in modern sensory neuroscience is that early sensory systems invert generative models of the environment to infer the hidden causes or latent variables that have produced sensory observations. Perhaps the simplest form of such inference is \emph{maximum a posteriori} inference, or MAP inference for short, in which the most likely configuration of latent variables given the sensory inputs is reported. The implementation of MAP inference in neurally plausible circuitry often requires all-to-all connectivity between the neurons involved in the computation. Given that the latent variables are often very high dimensional, this can imply single neurons being connected to millions of others, a requirement that is impossible to achieve in most biological circuits. Here we show how a MAP inference problem can be reformulated to employ sparse connectivity between the computational units. Our formulation is inspired by the vertebrate olfactory system, but is completely general and can be applied in any setting where such an inference problem is being solved.

We begin by describing the olfactory setting of the problem, and highlight the requirement of all-to-all connectivity. Then we show how the MAP inference problem can be solved using convex duality to yield a biologically plausible circuit. Noting that it too suffers from all-to-all connectivity, we then derive a solution inspried by the anatomy of the vertebrate olfactory that uses sparse connectivity.

\subsection{Sparse coding in olfaction}
We consider sparse coding \cite{olshausen_emergence_1996} as applied to olfaction \cite{koulakov_sparse_2011,grabska-barwinska_demixing_2013, tootoonian_dual_2014,grabska-barwinska_probabilistic_2017, kepple_deconstructing_2016}. Odors are modeled as high-dimensional, real valued latent variables $\xx\in\mathbb{\mathbb{{R}}}^{N}$ drawn from a factorized distribution
\begin{align}
  \tag{Odor model} p(\xx)=\prod_{i=1}^{N}p(x_i)=\frac{{1}}{Z}e^{-\phi(\xx)}, \quad \phi(\xx)= \beta\|\xx\|_{1}+\frac{{\gamma}}{2}\|\xx\|_{2}^{2} + \mathbb I(\xx \ge 0).
\end{align}
The first two terms of $\phi$ embody an elastic net prior \cite{kepple_deconstructing_2016,zou_regularization_2005} on molecular concentrations that models their observed sparsity in natural odors \cite{jouquand_sensory_2008}, while the last term enforces the non-negativity of molecular concentrations and is defined as $ \mathbb I(\xx \ge 0) = \sum_{i=1}^N \mathbb I(x_i \ge 0)$, where $\mathbb I(x_i \ge 0) =  0$ when $x_i \ge 0$ and $\infty$ otherwise. The animal observes these latents indirectly via low dimensional glomerular responses $\yy\in\mathbb{{R}}^{M}$, where $M\ll N$. Odors are transduced linearly into glomerular responses via the \emph{affinity matrix} $\AAA$, where $A_{ij}$ is the response of glomerulus $i$ to a unit concentration of molecule $j$. This results in a likelihoood $p(\yy|\xx)=\mathcal{{N}}(\yy;\AAA\xx,\sigma^{2}\II$), where $\sigma^{2}$ is the noise variance. As in \cite{tootoonian_dual_2014}, we assume that the olfactory system infers odors from glomerular inputs via MAP inference, i.e. by finding the vector $\xx_{\MAP}$ that minimizes the negative log posterior over odors given the inputs:
\begin{align}
  \tag{MAP inference} \xx_{\MAP}=\argmin_{\xx \in \bb R^N}\;\phi(\xx)+\frac{{1}}{2\sigma^{2}}\|\yy-\AAA\xx\|_{2}^{2}
\end{align}

\begin{figure}[h]
  \centering
  \includegraphics[width=\linewidth]{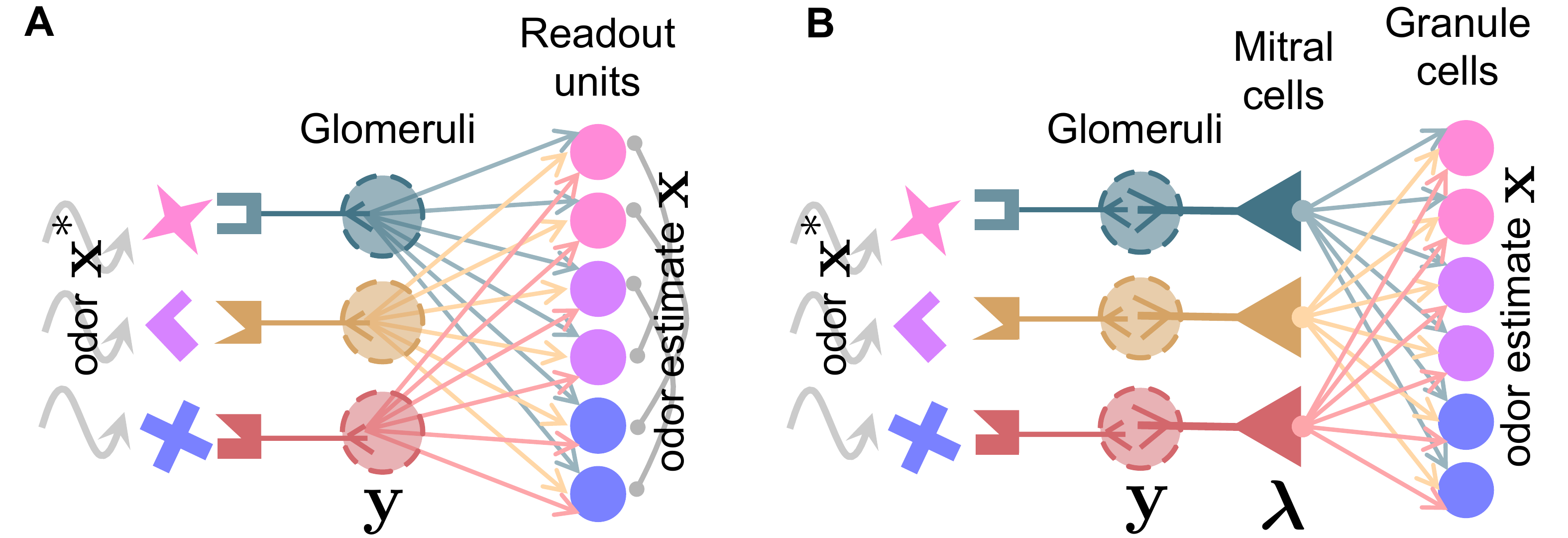}
  \caption{Two architectures for MAP inference requiring all-to-all connectivity in general. Arrows indicate excitatory connections, knobs indicate inhibitory connections. (A) Gradient descent architecture. All-to-all feedforward excitation is required from the glomeruli to the readout units, and all-to-all recurrent inhibition between the readout units. (B) Predictive coding architecture. All mitral cells excite all granule cells and are in turn inhibited by them. No direct interaction among granule cells is required. Both architectures yield the MAP solution at convergence.}
  \label{fig:all-to-all}
\end{figure}

A common approach to solving such problems is gradient descent \cite{olshausen_emergence_1996}, with dynamics in $\xx$:
\begin{align}
  \tag{Gradient descent}
  \tau \frac{d\xx}{dt} &= -\text{(leak)} + \frac{1}{\sigma^2}\AAA^T\yy - \frac{1}{\sigma^2}\AAA^T\AAA \xx,
  \end{align}
where we've absorbed the effects of the prior into the leak term for simplicity. These dynamics have a neural interpretation as feedforward excitation of the readout units $\xx$ by the glomeruli $\yy$ due to the $\AAA^T \yy$ term, and recurrent inhibition among the readout units due to the $-\AAA^T\AAA \xx$ term. This circuit is shown in Figure~\ref{fig:all-to-all}A.

Another circuit is motivated by noting that $\AAA^T\yy - \AAA^T \AAA \xx = \AAA^T(\yy - \AAA \xx)$. This suggests a predictive coding \cite{rao_predictive_1999} reformulation:
\begin{align}
  \tag{Predictive coding}
  \tau_{\text{fast}} \frac{d\rr}{dt} = -\text{(leak)} + \yy - \AAA \xx, \quad \tau_{\text{slow}} \frac{d\xx}{dt} = -\text{(leak)} + \frac{1}{\sigma^2}\AAA^T\rr.
  \end{align}
Here the new variable $\rr$ encodes the residual after explaining the glomerular activations $\yy$ with odor $\xx$. The neural interpretation of these dynamics is that the residual units $\rr$ receive feed-forward input from the glomeruli due to the $\yy$ term and feedback inhibition from the readout units due to the $-\AAA \xx$ term, while the readout units receive feedforward excitation from the residual units due to the $\AAA^T \rr$ term. This circuit is shown in Figure~\ref{fig:all-to-all}B.
\subsection{The problem of all-to-all connectivity}
Connectivity in the above circuits is determined by the affinity matrix $\AAA$. Given the combinatorial nature of receptor affinities \cite{nara_large-scale_2011}, $\AAA$ can be dense, i.e. have many non-zero values. This will result in correspondingly dense, even all-to-all connectivity. For example, the gradient descent architecture would require each glomerulus to connect to every readout unit, and for each readout unit to connect to every other. If we assume that the cells in the piriform cortex correspond to the readout units, this will require, in the case of the mouse olfactory bulb, that each glomerulus directly connect to millions of piriform cortical neurons, and for each cortical neuron to directly connect to millions of others. Such dense connectivity is clearly biologically implausible. The predictive coding circuit obviates the need for recurrent inhibition among the readout units, but still requires each residual unit to excite and receive feedback from millions of cortical neurons, which again is implausible. This problematic requirement of all-to-all connectivity is not limited to olfaction: the sparse coding formulation above is quite generic so that any system thought to implement it, such as the early visual system \cite{olshausen_sparse_1997}, is likely to face a similar problem. 
\section{Results}
To address the problem of all-to-all connectivity we will first show how MAP inference can be solved as a constrained optimization problem, resulting in a principled derivation of the predictive coding dynamics derived heuristically above. The resulting circuit also suffers from all-to-all connectivity. Taking inspiration from the anatomy of the olfactory bulb, we then show how the problem can be reformulated and solved using sparse connectivity.
\subsection{MAP inference as constrained optimization}
The MAP inference problem is a high-dimensional \emph{unconstrained} optimization problem, where we search over the full $N$-dimensional space of odors $\xx$. In \cite{tootoonian_dual_2014} the authors showed how a similar compressed-sensing problem can be solved in the lower-, $M$-dimensional space of observations by converting it to a low-dimensional \emph{constrained} optimization problem. Here we use similar methods to demonstrate how the MAP problem itself can be solved in the lower-dimensional space. We begin by introducing an auxiliary variable $\rr$, and reformulate the problem as constrained optimization:
\begin{align}
  \tag{MAP inference, constrained} \xx_{\MAP}, \rr_{\MAP}=\argmin_{\substack{\xx \in \bb R^N \\\rr \in \bb R^M}}\;\phi(\xx)+\frac{{1}}{2\sigma^{2}}\|\rr\|_{2}^{2} \quad \text{s.t.}\quad \rr = \yy - \AAA \xx.
\end{align}
The Lagrangian for this problem is
$$\Lag(\xx,\rr,\la)=\phi(\xx)+\frac{1}{2\sigma^2}\|\rr\|_2^2 + \la^T(\yy - \AAA \xx - \rr),$$
where $\la$ are the dual variables enforcing the constraint. The auxillary variable $\rr$ can be eliminated by extremizing $\Lag$ with respect to it:
$$\nabla_{\rr}\Lag = \frac{1}{\sigma^2}\rr - \la,\quad \nabla_{\rr}\Lag = 0 \implies \rr = \sigma^2 \la.$$
Plugging this value of $\rr$ into $\Lag$ we get
$$ \Lag(\xx,\la)=\phi(\xx)-\frac{1}{2}\sigma^2\|\la\|_2^2 + \la^T(\yy - \AAA \xx).$$
After a change of variables to $\la \leftarrow \sigma \la$ (which we justify below) we arrive at
\begin{align}
  \tag{MAP Lagrangian}\Lmap(\xx,\mc)=\phi(\xx)-\frac{1}{2}\|\mc\|_{2}^{2}+\frac{1}{\sigma}\mc^{T}(\yy-\AAA\xx).
\end{align}
Extermizing $\Lmap$ yields dynamics
\begin{align}
  \tag{Mitral cell firing rate relative to baseline}
  \tau_{mc} \frac{d\mc}{dt} &= - \mc + \frac{1}{\sigma}(\yy - \AAA \xx)\\
  \tag{Granule cell membrane voltage}
  \tau_{gc} \frac{d\vv}{dt} &= - \vv + \AAA^T\mc,\\
  \tag{Granule cell firing rate}
  \xx &= \frac{1}{\gamma \sigma}[\vv - \beta \sigma]_+,
\end{align}
where $[z]_+ = \text{max}(z,0)$ is the rectifying linear function. These dynamics can easily be shown to yield the MAP solution in the value of $\xx$ at convergence (see Supplementary Information). The identification of $\mc$ and $\xx$ with mitral and granule cells, respectively is natural as the dynamics indicate that (a) the $\mc$ variables are excited by the sensory input $\yy$ and inhibited by $\xx$, whereas (b) the much more numerous $\xx$ variables receive their sole excitation from the $\mc$ variables, and (c) the connectivity of the $\mc$ and $\xx$ variables is symmetric, reminiscent of the observed dendro-dendritic connections between mitral and granule cells \cite{shepherd_synaptic_2004}. The rescaling applied to $\mc$ is to keep mitral cell activity at convergence on the same order of magnitude as that of the receptor neurons, as qualitatively observed experimentally (compare for example \cite{shusterman_precise_2011} and \cite{duchamp-viret_odor_1999}): We assume without loss of generality that the elements of $\AAA$ and $\xx$ are scaled such that the elements of $\yy$ are $O(1)$ in magnitude. At convergence, $\la = \sigma^{-1}(\yy - \AAA \xx)$, and as we expect the elements of $\yy - \AAA \xx$ to be $O(\sigma)$ at convergence, this results in the elements of $\la$ being $O(1)$ in magnitude, as desired.

It may seem odd that the readout of the computation is in the activity of the granule cells, which not only do not project outside of the olfactory bulb, but lack axons entirely \cite{shepherd_synaptic_2004}. However, cortical neurons can read out the results of the computation by simply mirroring the dynamics of the granule cells:
\begin{align}
  \tag{Piriform cell membrane voltage}
  \tau_{pc} \frac{d\uu}{dt} &= - \uu + \AAA^T\mc,\\
  \tag{Piriform cell firing rate}
  \zz &= \frac{1}{\gamma \sigma}[\uu - \beta \sigma]_+,
\end{align}
In this circuit cortical neurons receive exactly the same mitral cell input as the granule cells and integrate it in exactly the same way (in fact, there is an implied 1-to-1 correspondence between granule cells and piriform cortical neurons) but are not required to provide feedback to the bulb. Thus, basic olfactory inference can be performed entirely within the bulb, with the concomitant increases in computational speed, and the results can be easily read out in the cortex. As cortical feedback to the bulb (in particular to the granule cells, as this model would suggest) does exist \cite{shepherd_synaptic_2004}, its role may be to incorporate higher level cognitive information and task contingencies into the inference computation. We leave the exploration of this hypothesis to future work.

These dynamics and their implied circuit are essentially the same as those of predictive coding described in the Introduction (Figure~\ref{fig:all-to-all}B), and hence suffer from the same problem of all-to-all connectivity. However, as we have derived our dynamics in a principled way from the original MAP inference problem, we can now elaborate them by taking inspiration from olfactory bulb anatomy to derive a circuit that can perform MAP inference but with sparse connectivity.

\subsection{Incorporating sister mitral cells  }
\begin{figure}[h]
  \centering
  \includegraphics[width=4in]{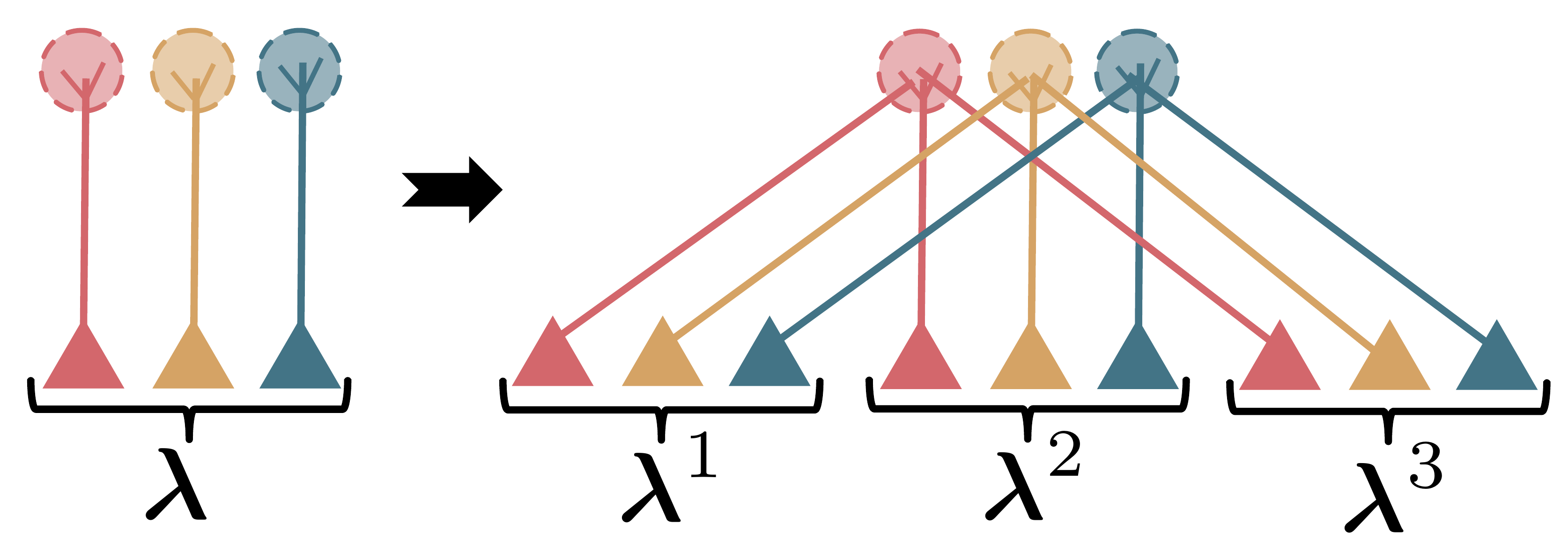}
  \caption{Sister mitral cells. In the vertebrate olfactory bulb, each glomerulus is sampled by not one but $\sim 25$ `sister' cells \cite{shepherd_synaptic_2004}. Here we've shown a setting with 3 sisters/glomerulus.}
  \label{fig:sister_mcs}
\end{figure}

The circuit derived above (Figure~\ref{fig:all-to-all}C) implies that each glomerulus is sampled by a single mitral cell. However, in vertebrates there are many more mitral cells than glomeruli, but each mitral cell samples a single glomerulus, so that each mitral cell has several dozen `sister' cells all of whom sample the same glomerulus \cite{shepherd_synaptic_2004}. This is shown schematically in Figure~\ref{fig:sister_mcs}. Although sister mitral cells receive the same receptor inputs their odor responses can vary, presumably due to differing interactions with the granule cell population \cite{dhawale_non-redundant_2010}. The computational role of the sister mitral cells has thus far remained unclear. Here we show that how they can be used to perform MAP inference but with sparse connectivity.

\begin{figure}[h]
  \centering
  \includegraphics[width=4in]{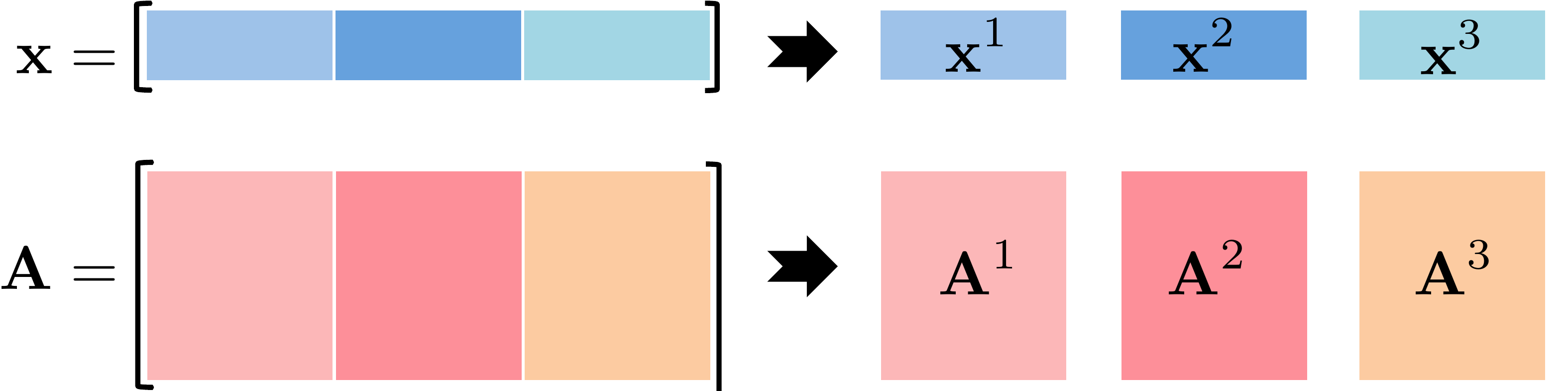}
  \caption{An example partitioning of the affinity matrix $\AAA$ and the odor vector $\xx$.}
  \label{fig:partition}
\end{figure}

We begin by noting the simple equalities
$$ \AAA \xx = \sum_{i=1}^n \AAA^i \xx^i, \quad \phi(\xx) = \sum_{i=1}^n \phi(\xx^i),$$
for any separable function $\phi$ (such as ours), and any  partitioning of the matrix $\AAA$ and corresponding partitioning of the vector $\xx$ into $n$ blocks. For example, if we partition $\AAA$ and $\xx$ into consecutive blocks, we'd have:
$$ \AAA = [\underbrace{A_{:,1},\dots,A_{:,N/n}}_{\AAA^1},\dots,\underbrace{A_{:,N-N/n+1},\dots,A_{:,N}}_{\AAA^n}], \quad  \xx = [\underbrace{x_1,\dots,x_{N/n}}_{\xx^1},\dots,\underbrace{x_{N-N/n+1},\dots,x_N}_{\xx^n}].$$
This partitioning is shown schematically in Figure~\ref{fig:partition}.

We can rewrite the Lagrangian $\Lmap$ in terms of this partitioning as
$$\Lmap(\xx, \la) = \Lmap(\{\xx^i\},\la)=-\frac{1}{2}\|\la\|_{2}^{2} + \frac{1}{\sigma}\la^T\yy + \sum_{i=1}^n \phi(\xx^i) -  \frac{1}{\sigma}\la^{T}\AAA^i \xx^i.$$
Note that although we've split $\AAA$ and $\xx$ into $n$ blocks, we're still using a single, shared $\la$ variable. Extremizing with resepect to the $\{\xx^i\}$ and a shared $\la$ would be an application of dual decomposition \cite{boyd_distributed_2011} to our problem. Instead, inspired by the presence of sister mitral cells, we reformulate the Lagrangian $\Lmap$ by assigning to each block its own set $\la^i$ of mitral cells, and introduce a corresponding set of variables $\mm^i$ to enforce the constraint $\la^i = \la$. This yields
\begin{align*}
  \Lsis(\{\xx^i\},\{\la^i\},\{\mm^i\},\la) = \sum_{i=1}^n \frac{1}{n\sigma }\la^{i,T}\yy + \phi(\xx^i) &- \frac{1}{2n} \|\la^i\|_2^2  - \frac{1}{\sigma}\la^{i,T}\AAA^i \xx^i\\
  &+ \mm^{i,T}(\la - \la^i) - \frac{1}{2}\|\la - \la^i\|_2^2.
\end{align*}
The additional term  $\frac{1}{2}\|\la - \la^i\|_2^2$ has been introduced because it does not alter the value of $\Lsis$ at the solution (since there $\la = \la^i$), while allowing us to eliminate $\la$ by setting $\nabla_{\la}\Lsis = 0$, yielding:
$$ \la =  \overline{\la} + \overline{\mm},\quad \overline{\la} = \frac{1}{n}\sum_{i=1}^n \la^i, \quad \overline{\mm} = \frac{1}{n}\sum_{i=1}^n \mm^i.$$
The values $\overline{\la}$ and $\overline \mm$ are averages computed over blocks, and are variables that would be available at the glomeruli. For example $\overline{\la}_i$ would be the average activity of all sister cells that innervate the $i$'th glomerulus.

As before, we derive dynamics by extermizing a Lagrangian, in this case $\Lsis$. As the $\{\mm^i\}$ are the dual variables of a constrained \emph{maximization} problem (that of maximizing $\Lsis$ with respect to $\{\la^i\}$), their dynamics minimize $\Lsis$:
$$ \frac{d\mm^i}{dt} \propto -\nabla_{\mm^i}\Lsis = \la^i - \la =  \la^i -  \overline{\la}  - \overline{\mm}  \implies \frac{d\overline{\mm}}{dt} \propto -\overline{\mm}.$$
Hence $\overline{\mm}$ decays to zero irrespective of the other variables, and in particular, if it starts at 0 it will remain there. In the following we will assume that this initial condition is met so that $\overline \mm = 0$ at all times, allowing us to eliminate it from the equations. The resulting dynamics that extremize $\Lsis$ are:
\begin{align*}
  \tag{Mitral cell activity relative to baseline}\tau_{mc} \frac{d\mc^i}{dt} &= -(1 + \frac{1}{n})\mc^i + \frac{1}{\sigma}\left(\frac{\yy}{n}  - \AAA^i \xx^i\right) +  \overline{\mc} - \mm^i\\
  \tag{Granule cell membrane voltage}\tau_{gc} \frac{d\vv^i}{dt} &= - \vv^i + \AAA^{i,T}\mc^i\\    
  \tag{Granule cell firing rate}\xx^i &= \frac{1}{\gamma \sigma }[\vv^i - \beta \sigma]_+\\  
  \tag{Periglomerular cell activity relative to baseline, no leak}\tau_{pg} \frac{d\mm^i}{dt} &=  \la^i - \overline{\la}
\end{align*}
We have identified the $\mm^i$ variables with olfactory bulb periglomerular cells because they inhibit the mitral cells and are in turn excited by them \cite{shepherd_synaptic_2004} and do not receive direct receptor input themselves, reminiscent of the Type II periglomerular cells of Kosaka and Kosaka \cite{kosaka_synaptic_2005}.

\begin{figure}[h]
  \centering
  \includegraphics[width=4in]{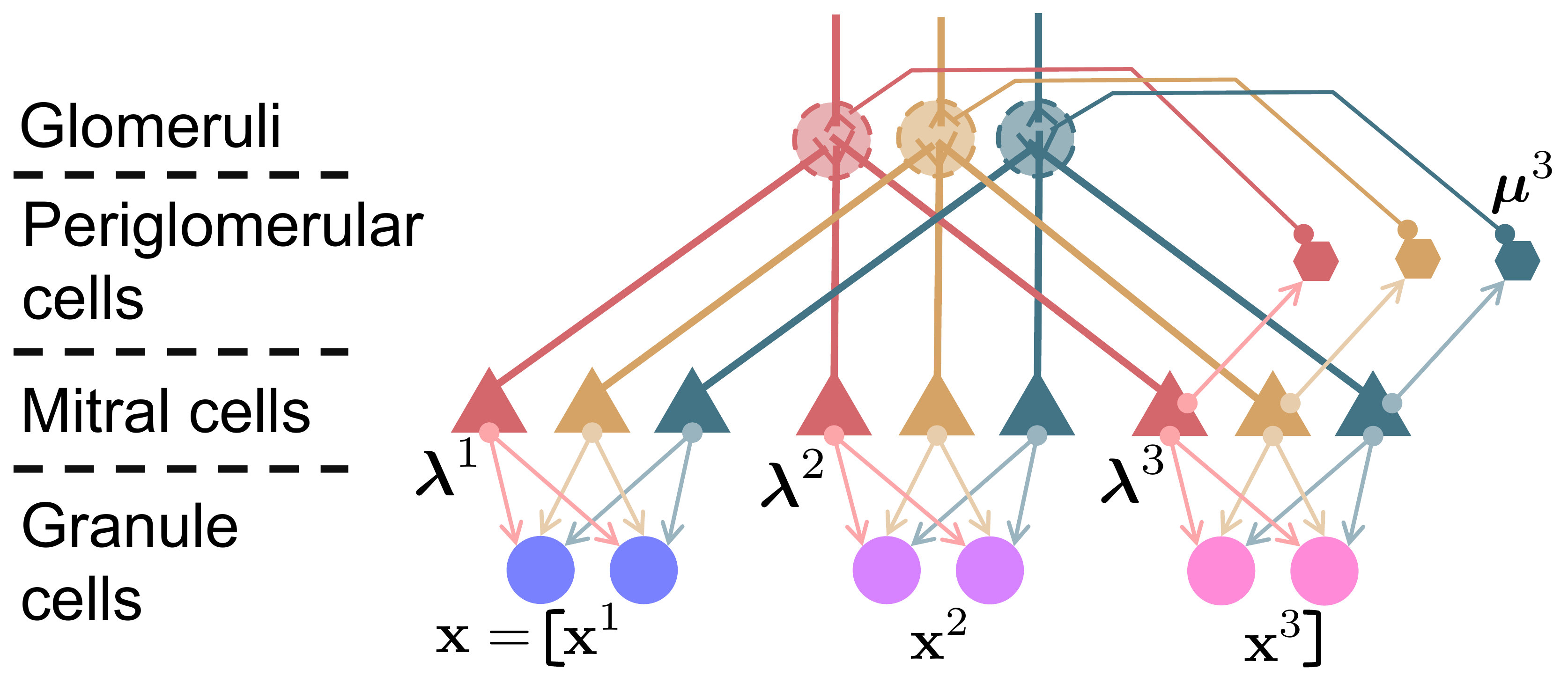}
  \caption{Inference circuit with sparse connectivity using sister mitral cells. Sister cells now only interact with the granule cells within their own block, reducing their connectivity by a factor of $n$. Information is shared between blocks at the glomeruli and through the periglomerular cells.}
  \label{fig:sparse_circuit}
\end{figure}

This circuit is shown schematically in Figure~\ref{fig:sparse_circuit}. Importantly, in this circuit each mitral cell interacts only with the granule cells within its block, reducing mitral-granule connectivity by a factor of $n$ (though the \emph{total} number of mitral-granule synapses has stayed the same due to the introduction $n$ sister mitral cells per glomerulus). The information from the other granule cells is delivered indirectly to each mitral cell via the influences of the glomerular average $\overline \la$ and periglomerular inhibition. 
\subsection{Leaky periglomerular cells via an approximate Lagrangian}
The dynamics above imply that that the periglomerular cells $\mm^i$ do not leak i.e. are perfect integrators, a property that is obviously at odds with biology. To introduce a leak term we first recall that $\mm^i$ dynamics minimize $\Lsis$. We then introduce an upper bound to $\Lsis$:
\begin{align*}
  \Lsiseps (\{\mm^i\},\dots)  = \Lsis(\{\mm^i\},\dots) + \sum_{i=1}^n \frac{1}{2}\|\la - \la^i\|_2^2 - \frac{1}{2(1 + \varepsilon)}\|\la - \la^i\|_2^2 + \frac{1}{2}\varepsilon \|\mm^i\|_2^2,
\end{align*}
where $\varepsilon \ge 0$ and we've suppressed the other arguments to the Lagrangians for clarity. The first two terms in the augmentation replace each $-\frac{1}{2}\|\la - \la^i\|_2^2$ term in $\Lsis$ with $-\frac{1}{2(1+\varepsilon)}\|\la - \la^i\|_2^2$, and the final term penalizes large values of $\mm^i$. The dynamics that extremize $\Lsiseps$ are the same as those that $\Lsis$ above, except for those of the mitral and periglomerular cells, which are modified to:
\begin{align*}
  \tau_{mc} \frac{d\mc^i}{dt} &= -(\frac{1}{1+\varepsilon} + \frac{1}{n})\mc^i + \frac{1}{\sigma}\left(\frac{\yy}{n}  - \AAA^i \xx^i\right) + \frac{\overline{\mc}}{1+\varepsilon} - \mm^i\\  
  \tau_{pg} \frac{d\mm^i}{dt} &= - \mm^i + \frac{1}{\varepsilon}(\la^i - \overline{\la})
\end{align*}
Note that now the periglomerular cells are endowed with a leak, as desired. Because the resulting dynamics no longer extremize $\Lsis$, the solution no longer matches the MAP solution exactly, and is in fact denser. To understand this effect (see Supplementary Information), note that at $\varepsilon = 0$, $\Lsiseps = \Lsis$, and the sister cells are `fully coupled' i.e. the $\mm^i$ variables are free to enforce the constraint $\la^i = \la$. The system then solves the MAP problem exactly by combining information from all blocks, yielding a sparse solution. As $\varepsilon \to \infty$ non-zero values of $\mm^i$ result in progressively higher values for the Lagrangian, forcing $\mm^i$ to zero in the limit. In this `fully decoupled' state each block attempts to explain its fraction $\yy/n$ of the input independently of the others using only its own subset $\AAA^i$ of the affinity matrix, reducing overcompleteness and resulting in denser representations. For the small values of $\varepsilon$ this can be counteracted by increasing the sparsity prior coefficient $\beta$.

\begin{figure}[h]
  \centering
  \includegraphics[width=\linewidth]{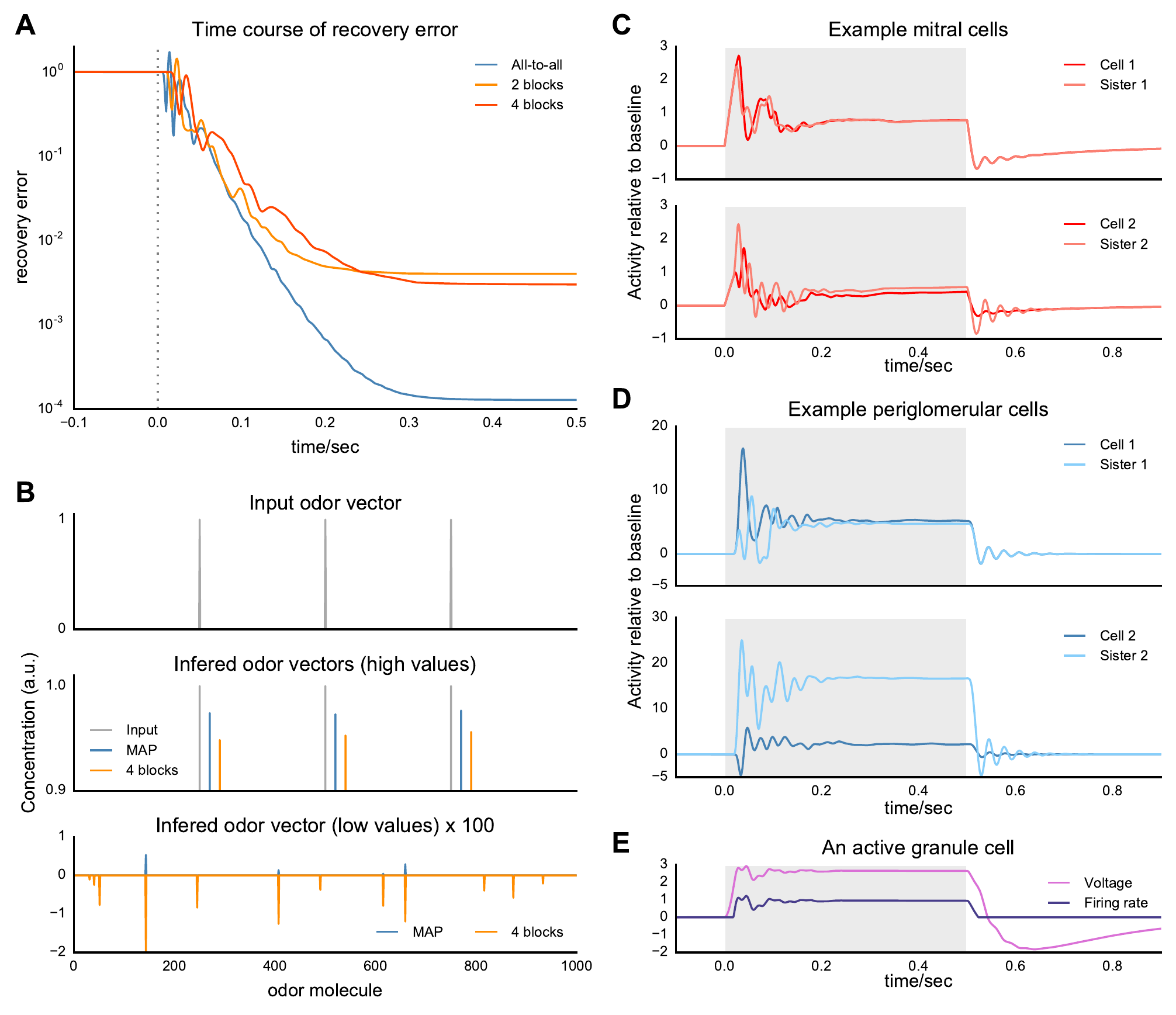}
  \caption{Performance. (A) Time course of recovery error for different circuits averaged over 40 random odors puffed for 500 ms at $t = 0$. Recovery error is mean squared error of granule cell activity relative to the MAP estimate, normalized by mean sum of squares of the MAP estimate. The all-to-all circuit essentially recovers the MAP solution; $n$-block circuits do so approximately as $\varepsilon>0$. Odors were sparse 1000-dimensional vectors ($N=1000$) with 3 randomly selected element set to 1. All-to-all circuit had 50 mitral cells ($M=50$); $n$-block circuits had $50n$ mitral cells and the corresponding periglomerular cells. Other parameters: $\beta = 100\;\text{(all-to-all)}, 150\;\text{(2-block)}, 170;\text{(4-block)}$;  $\gamma = 100$; $\sigma = 10^{-2}$ (but no noise was actually added above); $\varepsilon = 10^{-2}$; $\tau_{mc} = \tau_{pg} = \tau_{gc} = \text{50 ms}$. (B) Example recovery. The output of a circuit is the vector of granule cell activations immediately before odor offset. Top panel: actual odor presented. Bottom two panels: MAP estimate (blue) and output of the 4-block circuit (orange), zoomed in (and sign-inverted in the bottom panel) to values near 1 and 0, respectively, to highlight discrepancies between the MAP estimate and the circuit output, demonstrating good agreement. (C) Sister mitral cells: The time course of two mitral cells and one each of their sisters, showing that the activities of sister cells are similar but not identical. (D) The activity of the periglomerular cells paired to the mitral cells in (C). (E) The membrane voltage and firing rate of a granule cell strongly activated by the odor. Firing rate is stable by $\sim 200$ ms after odor onset, consistent with fast odor discrimination in rodents \cite{uchida_speed_2003}.}
  \label{fig:performance}
\end{figure}

Figure~\ref{fig:performance}A demonstrates the time course of the recovery error of the circuit in response to a 500 ms odor puff, as the number of blocks is varied, and averaged over 40 trials. Recovery error is defined as the mean sum of squares of the difference between the circuit's estimate and the MAP solution normalized by the mean sum of squares of the MAP solution. The all-to-all circuit is able to reduce this error to near zero (numerical precision) as it is performing MAP inference exactly. As the multi-block circuits use a non-zero value of $\varepsilon$ they are only approximating the MAP solution, but can still greatly reduce the recovery error when using an optimized setting of the sparsity parameter $\beta$, as described above. Figure~\ref{fig:performance}B shows the output of the 4-block circuit for a typical input odor, demonstrating its close approximation to the MAP solution. In Figure~\ref{fig:performance}C the dynamics of two different cells and their sisters from another block are shown, demonstrating that they are similar, but not identical, as experimentally observed \cite{dhawale_non-redundant_2010}, and Figure~\ref{fig:performance}D shows the activity of corresponding periglomerular cells. Finally, in Figure~\ref{fig:performance}E shows the membrane voltage and output firing rate of one of the active granule cells. Note that the firing rate has essentially stabilized by $\sim$200 ms after odor onset, broadly consistent with the fast olfactory discrimination times observed in rodents \cite{uchida_speed_2003}.

\section{Discussion}
Inspired by the sister mitral cells in the olfactory bulb, we have shown in this work how MAP inference, which often requires dense connectivity between computational units, can be reformulated in a principled way to yield a circuit with sparse connectivity, at the cost of introducing additional computational units. A key prediction of our model may appear to be that the mitral-granule cell connectome has block structure, in which granule cells only communicate with the mitral cells in their block and vise versa. As we show in the Supplemental Information, a simple generalization of our model shows that MAP solution can be found with mitral-granule cell connectivity that does not have the block structure we have assumed here (though equally sparse). This generalization also accommodates the the experimentally observed random sampling of glomeruli by mitral cells \cite{imai_construction_2014}, in addition to the ordered one presented above where exactly $n$ sister cells sample each glomerulus. 

Previous work in several groups has addressed sparse coding in olfaction \cite{koulakov_sparse_2011,grabska-barwinska_demixing_2013, tootoonian_dual_2014, grabska-barwinska_probabilistic_2017, kepple_deconstructing_2016}. Our work extends that of \cite{tootoonian_dual_2014} in insects by showing how the MAP problem itself can be solved, rather than the related compressed sensing problem addressed in that paper. In our work we propose that olfactory bulb granule cells encode odor represenations, similar to \cite{koulakov_sparse_2011}. The authors there assumed a random mitral-granule connectome, resulting in `incomplete' odor represenations because granule cell firing rates are positive. In this work we assume that the connectome is set to its `correct' value determined by the affinity matrix $\AAA$, obviating the need for negative rates and resulting in `complete' representations. Even with such complete representations, mitral cell activity is not negligible, and allows for simple and exact readout of the infered odor concentrations in downstream cortical areas. Furthermore, previous work \cite{tootoonian_dual_2014} has shown, albeit in a limited setting, that the correct value of the connectome can be learned via biologically plausible learning mechanisms. We expect that to be the case here, though we leave that determination to future work. The authors in \cite{grabska-barwinska_demixing_2013, grabska-barwinska_probabilistic_2017} propose a model in which the olfactory bulb and cortex interact to infer odorant concentrations while retaining uncertainty information, rather than just providing point estimates as in MAP inference. The authors in \cite{kepple_deconstructing_2016} propose a bulbar-cortical circuit that represents odors based on `primacy', the relative strengths of the strongest receptor responses, automatically endowing the system with the concentration invariance likely to be important in olfactory computation. We've shown that the MAP computation can be performed entirely within the bulb while allowing for easy and exact cortical readout and without the need for cortical feedback, retaining odor information and allowing downstream areas to perform concentration invariance and primacy computations, as needed. Extending our methods to provide uncertainity information is an important task that we leave  to future work.
\bibliographystyle{unsrt}
\clearpage
\pagebreak
\bibliography{ms}
\pagebreak
\section*{Supplementary Information}
\subsection*{Generalizing the model}
\begin{figure}[h]
  \centering
  \includegraphics[width=\linewidth]{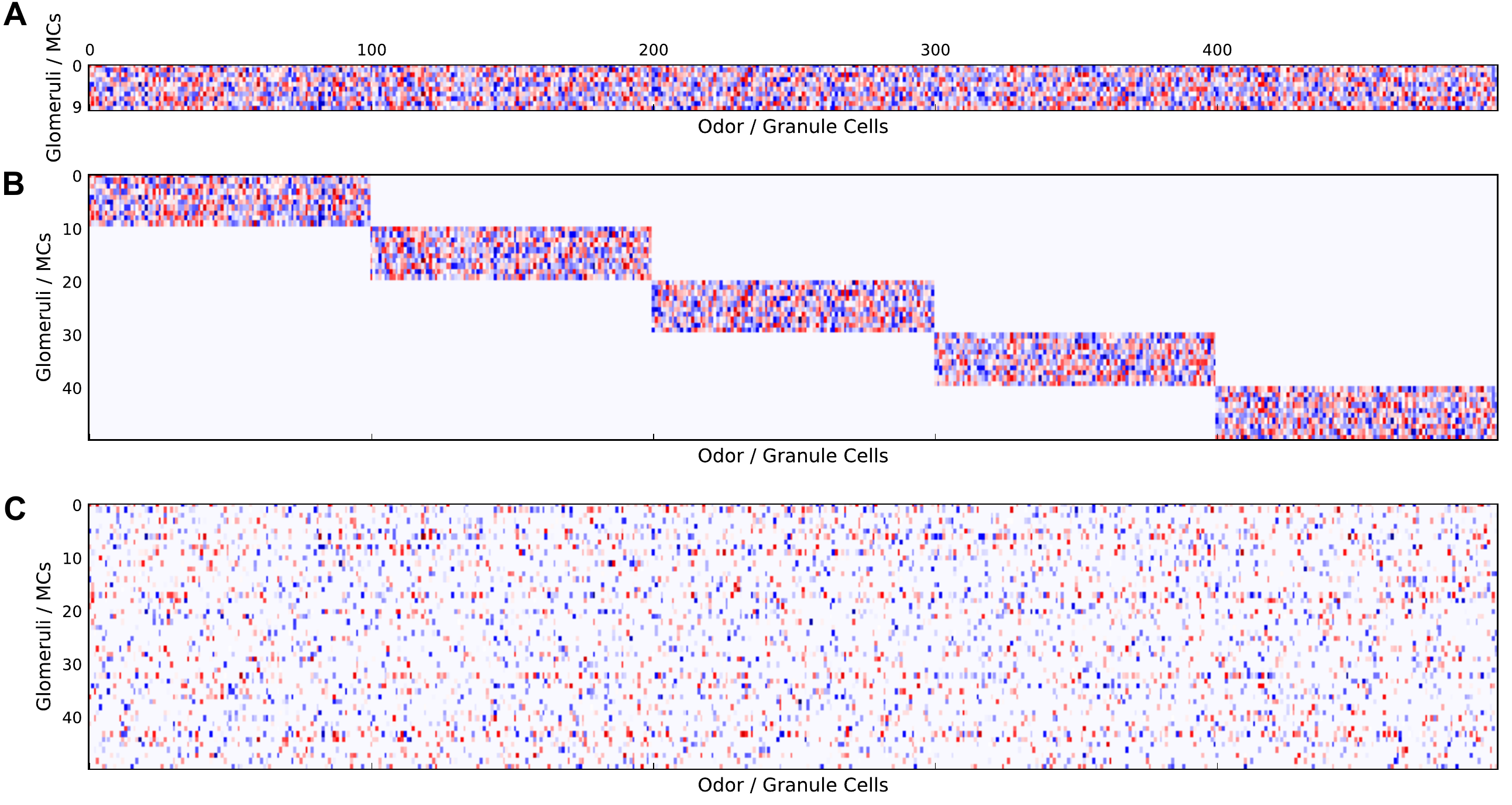}
  \caption{Sparsifying connectivity. (A) The original all-to-all affinity matrix $\AAA$. (B) The affinity matrix sparsified by being split into blocks, as considered in the main text. (C) A learned affinity matrix with the same number of non-zero elements as that in (B), but without block structure. Both the matrices $\WW$ in panels (B) and (C) satisfy the third matrix condition $\VV^T\DD^T \WW = \AAA$ for $\VV = \bb 1_n \otimes \II_M$ and $\DD = \II_{nM}$, demonstrating that a block diagonal mitral-granule connectome is not required for circuit dynamics to yield the MAP solution.}
  \label{fig:connectivity}
\end{figure}

Our model as formulated in the main text predicts that granule cells interact only with the mitral cells within their blocks. This predicts a block diagonal structure in the mitral-granule cell connectome, such as in Figure~\ref{fig:connectivity}B. However, this is not the only possible solution. To determine the set of all possible solutions we extend the derivations in the main text to a slightly more general setting. This generalization will also allow us to deal with the biologically observed random sampling of glomeruli by mitral cells in the following section.

Instead of considering the sister mitral cells separately as $\{\la^i\}$, we can stack them into one large vector $\La$, similarly stack the periglomerular celsl $\{\mm^i\}$ into $\Mu$ and consider a generalized Lagrangian
$$ \Lsiseps(\xx,\La, \la, \mm)= \phi(\xx)-\frac{1}{2}\|\FF \La\|_{2}^{2}+\frac{1}{\sigma}{\La}^{T}(\GG \VV \yy-\WW \xx) + \mm^T (\DD \VV \la - \La) - \frac{\|\DD \VV \la - \La\|_2^2}{2(1+\varepsilon)} + \frac{\varepsilon}{2}\|\mm\|_2^2.$$
Here $\bb F = \text{diag}(f_1,\dots,f_T)$ and $\bb G= \text{diag}(g_1,\dots,g_T)$, where $T$ is the total number of mitral cells. The binary matrix $\VV$ indicates the glomeruli sampled by each mitral cell. As each mitral cell samples exactly one glomerulus, each of the rows of $\VV$ contain just one non-zero element, rendering $\VV^T\VV$ orthogonal (though not orthonormal). $\GG$ is the gain each mitral cell applies its glomerular input, $\FF$ can modify the leak time constant of each mitral cell, and $\WW$ is the mitral-granule connectome. The relationship between $\La$ and $\la$ at convergence must satisfy $\La = \DD \VV \la$, to mirror the sampling of glomeruli by mitral cells, where we've included the diagonal matrix $\DD = \diag(\{d_i\})$ to allow for cell-specific gain. We can then ask what conditions these matrices ensure that when $\La = \DD \VV \la$, $\Lsis^0 = \Lmap$. Plugging $\DD \VV \la$ in for $\La$, we have
$$\Lsis^0(\xx,\La, \la, \mm) = \phi(\xx)-\frac{1}{2}\la^T \VV^T \DD^T \FF^T\FF \DD \VV \la +\frac{1}{\sigma}\la^T \VV^T \DD^T(\GG \VV \yy-\WW \xx).$$
Then by inspection, if the following conditions
$$ (1)\; \VV^T \DD^T \FF^T \FF \DD \VV = \bb I_M, \quad (2)\; \VV^T \DD^T \GG \VV = \bb I_M, \quad\text{and}\quad (3)\; \VV^T \DD^T \WW = \AAA$$ are met,
$$ \Lsis^0(\xx,\La, \la, \mm) = \phi(\xx)-\frac{1}{2}\|\la\|_2^2 +\frac{1}{\sigma}\la^T (\yy-\AAA \xx) = \Lmap.$$
Hence, extremizing $\Lsis^0$ will yield the MAP solution at convergence (see below for a direct derivation). $\Lsis$ considered in the main text corresponds to
$$\VV = \bb 1_n \otimes \II_M, \quad \FF^T\FF = \GG = \frac{1}{n}\II_{nM}, \quad \DD = \II_{nM},$$
and $\AAA$ (e.g. Figure~\ref{fig:connectivity}A) partitioned to yield a block-diagonal mitral-granule  (Figure~\ref{fig:connectivity}B). This setting of the matrices satisfies the conditions above, guaranteeing that extremizing $\Lsis$ in the text yields the MAP solution.

Note that the third condition above implies that any $\bb W$ that satisfies $\bb V^T \bb W = \bb A$ will result in the extremization of $\Lag_0$ and yield the MAP solution. Although the block structured connectome in Figure~\ref{fig:connectivity}B satisfies this conditions, so too does e.g. the connectome in Figure~\ref{fig:connectivity}C. This latter connectome was generated by performing a modified $\ell_0$ minimization on a connectivity matrix $\WW$, subject to the third constraint above. The result has the same sparseness as the matrix in Figure~\ref{fig:connectivity}B, but without the block structure. Thus a block-structured mitral-granule connectome is not the only sparsity pattern that solves the MAP solution, and given that the biological connectome is likely a result of learning, the experimentally observed connectome is more likely to resemble that in Figure~\ref{fig:connectivity}C, and to lack block structure. 

\subsubsection*{Random sampling of glomeruli}
In this work we've assumed that each glomerulus is sampled by exactly the same number $n$ sister mitral cells. In biological fact, glomeruli are sampled randomly by mitral cells, but subject to the constraint that each mitral cell samples exactly one glomerulus. As long as each glomerulus is sampled at least once, our generalized framework in the previous section can accommodate this situation. The random sampling of glomeruli can be modeled as each row of the matrix $\VV$ having one \emph{randomly selected} element set to 1. Setting $\DD = \II_{nM}$, condition (2) then implies
$$ \VV^T \GG \VV = \II_M \implies \VV^T \GG \VV \bb 1 = \bb I_M \bb 1 \implies \VV^T \GG \bb 1 = \bb 1 \implies \VV^T \bb g = \bb 1,$$
where $\bb g$ is the vector of diagonal elements of $\GG$, and we've used the fact that $\VV \bb 1 = \bb 1$. As $\VV^T$ has more columns than rows, the last equation is under-determined, so we can take the solution with least Euclidean norm by assuming that $\bb g$ is in the range of $\VV$ i.e. $\bb g = \VV \bb g'$. We then have
$$ \VV^T \bb g= \bb 1 \implies \VV^T \VV \bb g' = \bb 1 \implies \bb g' = [n_1^{-1},\dots,n_M^{-1}]^T \implies \GG = \text{diag}(\{n_i^{-1}\}),$$
where $n_i$ is the number of sisters that mitral cell $i$ has. What this value of $\GG$ implies in terms of the circuit is that glomerular activation is split evenly among the innervating sister cells, so that each receives $1/n_i$ of the excitation. This process would occur naturally as the result of the neurotransmitter released by receptor neurons in the glomerulus being distributed approximately evenly among the innervating mitral cell dendrites. As for the remaining variables, $\bb F$ can then be set to $ \bb F = \text{diag}(\{n_i^{-1/2}\})$, and $\WW$ can be chosen arbitrarily as long as it satisfies $\VV^T \WW = \AAA$.
\subsubsection*{Generalizing the constraint on sister-cells}
As a final generalization, we consider the desired mapping between sister mitral cells $\La$ and the original mitral cell vector $\la$. In the main text, we've required that $\La = \VV \la$ at convergence but we can generalize this to  $\La = \DD \VV \la$, where $\DD = \diag(\{d_i\})$ can be interpreted as a gain applied to the constraint on each sister mitral cell. We can then use condition (2) to solve for $\GG$:
$$ \VV^T \DD^T \GG \VV = \II_M \implies \VV^T \DD^T \GG \VV \bb 1 = \VV^T  (\bb {d g}) = \bb 1,$$
where $\bb {dg} = [\{d_i g_i\}]^T$. As before, we can then assume that $\bb {dg}$ is in the range of $\VV$, which like before yields
$$ \VV^T (\bb {dg})= \bb 1 \implies \VV^T \VV (\bb{dg})' = \bb 1 \implies (\bb {dg})' = [n_1^{-1},\dots,n_M^{-1}]^T \implies \GG = \text{diag}(\{(d_i n_i)^{-1}\}).$$
To solve for $\FF$ we can then note that if condition 2 is satisfied, then condition 1 can be satisfied by setting $\FF^T\FF \DD = \GG$. We then have
$$ \diag(\{f_i^2d_i\}) = \diag(\{(d_i n_i)^{-1}\}) \implies f_i^2 d_i = (d_i n_i)^{-1} \implies f_i^2 = d_i^{-2}n_i^{-1} \implies f_i = (d_i \sqrt{n_i})^{-1},$$
so that $\FF = \diag(\{d_i^{-1}n_i^{-1/2}\}),$
showing that our framework can accommodate this setting as well.
\subsection*{Derivation of the generalized dynamics}
We will derive the dynamics for $\Lsiseps$ in the generalized setting introduced above. Dynamics for our variables are determined by extremizing this Lagrangian. We minimize with respect to $\xx$ as it's the primal variable, maximize with respect to $\la$ and $\La$ because they are dual variables, and minimize with respect to $\mm$ as it is the dual variable for the maximization with respect to $\la$ and $\La$.

We first eliminate $\la$ from the dynamics by setting its gradient to zero. The gradient is
$$\nabla_{\la} \Lsiseps  = \VV^T\DD^T \mm + \frac{1}{1 + \varepsilon}(\VV^T \DD^T \La - \VV^T \DD^T\DD \VV \la).$$
Setting it to zero yields
$$\VV^T \DD^T \DD \VV \la = \VV^T \DD^T( \La - (1 + \varepsilon) \mm) \implies \la =  (\VV^T \DD^T \DD \VV )^{-1}\VV^T \DD^T(\La + (1 + \varepsilon) \mm).$$
Thus $ \la$ are the coefficients of the least-squares projection of the $N$-dimensional variable $ \La + (1 + \varepsilon) \mm$ into the $M$-dimensional span of $\DD \VV$. Then
$$ \DD \VV \la = \DD \VV (\VV^T \DD^T \DD \VV )^{-1}\VV^T \DD^T ( \La + (1 + \varepsilon) \mm) = \PP (\La + (1 + \varepsilon) \mm)$$
is the projection of $\La + (1 + \varepsilon) \mm$ into the span of $\DD \VV$, and $\PP = \DD \VV (\VV^T \DD^T \DD \VV )^{-1}\VV^T \DD^T$ is the projection matrix.

The dynamics maximize $\Lsiseps$ with respect to $\La$:
\begin{align*}
  \dot{\La} \propto \nabla_{\La}\Lsis &= -\FF^T \FF \La + \frac{1}{\sigma}(\GG \VV \yy - \WW \xx) -  \mm + \frac{1}{1 + \varepsilon}(\DD \VV \la - \La)\\
  &= -\left[\FF^T \FF + \frac{1}{1+\varepsilon}\II\right] \La + \frac{1}{\sigma}(\GG \VV \yy - \WW \xx) - \mm  + \frac{1}{1+\varepsilon}\PP (\La + (1 + \varepsilon)\mm).
\end{align*}
The dynamics of $\mm$ minimize $\Lsiseps$ with respect to it:
\begin{align*}
  \dot \mm \propto -\nabla_{\mm} \Lsiseps &=  -\DD \VV \la + \La - \varepsilon \mm =  \La - \PP ( \La + (1 + \varepsilon)\mm) - \varepsilon \mm =  (\II - \PP)  \La - (1 + \varepsilon)\PP \mm - \varepsilon \mm.
\end{align*}
Thus we can decompose $\mm$ into orthogonal components $\PP \mm$ and $(\II - \PP) \mm$, with dynamics
\begin{align*}
  \PP \dot\mm \propto  -\PP \mm, \quad  (\II - \PP) \dot \mm \propto (\II - \PP) (\La - \varepsilon \mm).
\end{align*}
These imply that the $\PP \mm$ component decays to zero, and in particular that if it starts at zero, it will remain there. Therefore we will require that this initial condition holds so that $\PP \mm = 0$ for all $t$, and simplify the $\mm$ dynamics to 
$$ \dot \mm \propto -\mm + \frac{1}{\varepsilon} (\II - \PP)\La.$$
The dynamics for $\vv$ and $\xx$ remain unchanged, with $\vv$ integrating the input from $\La$ and $\xx$ applying a rectifying nonlinearity. If we identify the activity of the mitral cells with $\La$, the dynamics for the full system can then be defined as
\begin{align*}
  \tau_{mc}\frac{d\Mc}{dt} &= -(\FF^T\FF + \frac{1}{1+\varepsilon}\II) \Mc + \frac{1}{\sigma}(\GG \VV \yy - \WW \xx) + \frac{1}{1+\varepsilon}\PP \Mc - \mm\\
  \tau_{pg}\frac{d\mm}{dt} &= -\mm + \frac{1}{\varepsilon}(\Mc - \PP\Mc)\\
  \tau_{gc}\frac{d \vv}{dt} &= -\vv + \WW^T \Mc \\
  \xx &= \frac{1}{\gamma \sigma}[\vv - \beta \sigma]_+
\end{align*}
\subsection*{The MAP solution is the stationary point of the dynamics}
The MAP problem is to find the $\xx$ that minimizes the log posterior $\phi(\xx) + \frac{1}{2\sigma^2}\|\yy - \AAA\xx\|_2^2$. This $\xx$ satisfies
$$0 \in \partial \phi(\xx) -\frac{1}{\sigma^2}\AAA^T (\yy - \AAA \xx) \implies \xx = \frac{1}{\gamma \sigma^2}[\AAA^T (\yy - \AAA \xx) - \beta \sigma^2]_+,$$
where $\partial \phi(\xx)$ is the subgradient of $\phi$.

We will first show that the dynamics that extremize $\Lmap$ yield an $\xx$ variable that satisfies this relation. We will then show that the same is true for the generalized dynamics described above.

Dynamics that extremize $\Lmap$ are
\begin{align*}
  \tau_{\lambda} \frac{d\mc}{dt} &= - \mc + \frac{1}{\sigma}(\yy - \AAA \xx)\\
  \tau_{v} \frac{d\vv}{dt} &= - \vv + \AAA^T\mc,\\
  \xx &= \frac{1}{\gamma \sigma}[\vv - \beta \sigma]_+,
\end{align*}
At convergence, we have
$$ \frac{d\mc}{dt} = 0 \implies \la = \frac{1}{\sigma}(\yy - \AAA \xx).$$
$$ \frac{d\vv}{dt} = 0 \implies \vv = \AAA^T \la = \frac{1}{\sigma}\AAA^T(\yy - \AAA \xx).$$
$$ \xx = \frac{1}{\gamma \sigma}[\vv - \beta \sigma]_+ = \frac{1}{\gamma \sigma}[\frac{1}{\sigma}\AAA^T(\yy - \AAA \xx)  - \beta \sigma]_+.$$
Finally, using the fact that $[ax]_+ = a[x]_+$ for $a>0$ we have
$$ \xx = \frac{1}{\gamma \sigma^2}[\AAA^T(\yy - \AAA \xx)  - \beta \sigma^2]_+,$$
as desired.

We will now show that when the generalized dynamics above for $\varepsilon = 0$ converge, the $\xx$ variable satisfies this relation. We've assumed that $\PP \mm = 0$ as the dynamics will drive it there if it does not initially start at zero. Multiplying both sides with $\VV^T \DD^T$ and using the fact that $\VV^T \DD^T \PP = \VV^T \DD^T$, we have
$$ \PP \mm = 0 \implies \VV^T\DD^T \PP \mm = \VV^T \DD^T \mm = 0.$$
At convergence $\dot \mm = 0$ which means $\PP \Mc = \Mc$ which implies that $\Mc$ is in the range of $\DD \VV$, i.e. there exists $\la$ such that $\DD \VV\la = \Mc$. Also $\dot \Mc = 0$, so we have, after substituting $\Mc$ for $\PP \Mc$, $$\FF^T\FF \Mc = \sigma^{-1}(\GG \VV \yy - \WW \xx) - \mm.$$
Substituting $\DD \VV \la$ for $\Mc$, and multiplying both sides by $\VV^T \DD^T$, we have
$$  \VV^T \DD^T \FF^T \FF \DD \VV \la = \sigma^{-1}(\VV^T\DD^T\GG \VV \yy - \VV^T\DD^T \WW \xx)  - \VV^T \DD^T \mm.$$
Substituting in the three matrix constraints and the fact that at convergence $\VV^T \DD^T \mm = 0$, we have $$\la = \frac{1}{\sigma}(\yy - \AAA \xx).$$
Finally, $$\dot \vv = 0 \implies \vv = \WW^T \Mc =  \WW^T \DD \VV  \la =  \AAA^T \la,$$
so
\begin{align*}
  \xx = \frac{1}{\gamma \sigma}[\AAA^T \la - \beta \sigma]_+ = \frac{1}{\gamma \sigma}[\frac{1}{\sigma}\AAA^T (\yy - \AAA \xx)  - \beta \sigma]_+ = \frac{1}{\gamma \sigma^2 }[\AAA^T(\yy - \AAA \xx) - \beta \sigma^2]_+,
\end{align*}
as desired. Note that as $\Lolf$ is a particular case of $\Lsis$ in which $\Mc = \la$ and $\DD = \FF = \GG = \VV = \II_M$, so that $\PP\Mc = \Mc$ and the various matrix constraints are met, the fact that the generalized dynamics arrive at the MAP solution automatically guarantees that the dynamics extremizing $\Lolf$ also do.

\subsection*{Understanding the behaviour of the uncoupled circuit}
The generalized Lagrangian that accomodates leaky periglomerular cells is
$$ \Lsiseps(\xx,\La, \la, \mm)= \phi(\xx)-\frac{1}{2}\|\FF \La\|_{2}^{2}+\frac{1}{\sigma}{\La}^{T}(\GG \VV \yy-\WW \xx) + \mm^T (\DD \VV \la - \La) - \frac{1}{2(1+\varepsilon)}\|\DD \VV \la - \La\|_2^2 + \frac{1}{2}\varepsilon\|\mm\|_2^2.$$
Changing the $\varepsilon$ parameter allows us to vary the dynamics of the system a`fully coupled' state at $\varepsilon = 0$, to its `fully uncoupled' state as $\varepsilon \to \infty$. The fully coupled state $\Lsis^0$ is equivalent to $\Lsis$, in which the variables $\mm$ are free to enforce the constraint $\Mc = \DD \VV \la$, thus coupling the $\Mc$ variables and solving the MAP solution exactly, as we've shown above. In the fully uncoupled limit, any non-zero value of $\mm$ incurs infinite loss, clamping its value at 0. This implies that the $\Mc$ are no longer required to satisfy the $\Mc = \DD \VV \la$ constraint, hence our term `fully uncoupled' for this state of the circuit. The Lagrangian reduces to
$$ \Lsis^\infty(\xx,\La)= \phi(\xx)-\frac{1}{2}\|\FF \La\|_{2}^{2}+\frac{1}{\sigma}{\La}^{T}(\GG \VV \yy-\WW \xx),$$
which by inspection is just a larger version of the original MAP Lagrangian.

The behaviour of the fully uncoupled state is easiest to understand in the simple $n$-block setting, where each of the glomeruli is sampled evenly by the mitral cells and the $\AAA$ matrix is partitioned evenly among the blocks. This corresponds to a setting of
$$ \VV = \mathbf{1}_n \otimes \II_M, \quad \FF^T\FF = \GG = n^{-1}\II_{nM}, \quad \DD = \II_{nM}.$$
The Lagrangian then reduces to
\begin{align*}
  \Lsis^\infty(\xx,\La) &= \phi(\xx)-\frac{1}{2n }\|\La\|_{2}^{2}+\frac{1}{\sigma}{\La}^{T}(\frac{1}{n}\VV \yy -\WW \xx)\\
  &= \sum_{i=1}^n \phi(\xx^i)-\frac{1}{2n}\|\La^i\|_{2}^{2}+\frac{1}{\sigma}\La^{i,T}(\frac{\yy}{n}-\AAA^i \xx^i),\\
  &= \sum_{i=1}^n \ell(\xx^i, \La^i).
\end{align*}
Hence the Lagrangian is just the sum of $n$ terms that can be extremized independently, emphasizing the `fully uncoupled' nature of this state. To understand the nature of the solutions in this state, we note the similarity of each of the $\ell(\xx^i, \La^i)$ terms to the MAP Lagrangian $\Lmap$.
We have
\begin{align*}
  \ell(\xx^i, \La^i) &= \phi(\xx^i) - \frac{1}{2n}\|\La^i\|_{2}^{2}+\frac{1}{\sigma}\La^{i,T}(\frac{\yy}{n}-\AAA^i \xx^i).
\end{align*}
Rescaling $\La^i \leftarrow \La^i/\sqrt{n}$, we get  
$$  \ell(\xx^i, \La^i) =  \phi(\xx^i) - \frac{1}{2}\|\La^i\|_{2}^{2}+\frac{\sqrt{n}}{\sigma} \La^{i,T}(\frac{\yy}{n}-\AAA^i \xx^i) $$
We recognize this as a MAP Lagrangian, thus revealing that extremizing $\ell(\xx^i, \La^i)$ is equivalent to MAP inference but with input signal and noise variance scaled by $1/n$, and limited to $\AAA^i$ and the corresponding latents. Thus in the fully uncoupled regime each block attempts to explain its fraction $\yy/n$ of the input independently of the other blocks by solving its own MAP inference problem using only its own partition $\AAA^i$ of the affinity matrix, resulting in denser representations due to a reduction in overcompleteness.
\end{document}